# Photolysis of pure solid $O_3$ and $O_2$ films at 193 nm

U. Raut, M.J. Loeffler, M. Famá and R.A. Baragiola

University of Virginia, Laboratory of Atomic and Surface Physics

Charlottesville, VA 22904.

raut@virginia.edu, mjl8r@virginia.edu, maf7e@virginia.edu, raul@virginia.edu

## Abstract

We studied quantitatively the photochemistry of solid $O_3$ and $O_2$ films at 193 nm and 22 K with infrared spectroscopy and microgravimetry. Photolysis of pure ozone destroyed $O_3$, but a small amount of ozone remained in the film at high fluence. Photolysis of pure $O_2$ produced $O_3$ in an amount that increased with photon fluence to a stationary level. For both $O_2$ and $O_3$ films, the $O_3$:$O_2$ ratio at saturation is 0.03, nearly 10-30 times larger than those obtained in gas phase photolysis. This enhancement is attributed to the increased photodissociation of $O_2$ due to photoabsorption by $O_2$ dimers, a process significant at solid state densities. We obtain initial quantum yield for ozone synthesis from solid oxygen, $\Phi(O_3) = 0.18$ and for destruction of ozone and oxygen in their parent solids, $\Phi(-O_3) = 1.7$ and $\Phi(-O_2) = 0.28$. Combined with known photoabsorption cross sections, we estimate probabilities for geminate recombination of 0.15 for $O_3$ fragments and 0.90 for oxygen atoms from $O_2$ dissociation. Using a single parameter kinetic model, we deduce the ratio of reaction cross sections for an O atom with $O_2$ vs. $O_3$ to be 0.10. This ratio is the same for both $O_2$ and $O_3$ films, even though the distribution of kinetic and internal energy of the photofragments is very different. This finding suggests efficient energy relaxation of photofragments in the solid occur prior to their reactions with other species.

## 1. Introduction

For more than a century, photoabsorption of UV light by ozone has received great attention because of its importance in the protection to life by the ozone layer in the outer atmosphere and in the greenhouse effect and air pollution at ground level. In the last few decades, studies of the photolysis of solid ozone have been made to understand specific solid state effects[1,2] and to explain the observation of condensed ozone on the surfaces of cold satellites[3] of the outer planets. The proposal that radiolysis of $H_2O$ and $O_2$ ices is the source of the condensed ozone in planetary ices has been tested in experiments where thin films were irradiated with energetic particles or photons. Bahr et al found[4] that radiolysis of pure water ice does not produce any significant amounts of ozone but Teolis et al[5] discovered that ozone can indeed be produced if water was deposited during irradiation, which enhanced accumulation of radiolytic oxygen. In contrast, ozone is readily produced when $O_2$ [Ref. [6,7]] and $O_2 + H_2O$ films [Ref. [8,9]] are irradiated with energetic particles ($H^+$, $e^-$) and ultraviolet photons.[10] In addition to planetary applications, there are predictions by astrochemical models that condensed $O_2$ and $O_3$ should be present in the icy grain mantles in dense molecular clouds.[11-13] The presence of photosensitive molecules like $O_3$ on interstellar grains, which can yield reactive fragments by energetic radiation, could be crucial to chemistry occurring in dark molecular clouds due to a background ultraviolet flux from the decay of hydrogen excited by cosmic rays.

The vast majority of the work on photolysis of oxygen and ozone has been done in the gas phase, motivated by the need to understand the atmospheric ozone problem. Such studies have revealed the existence of multiple photolysis channels with different energy thresholds and involving many excited atomic and molecular states.[14] By comparison, the photolysis of ozone and oxygen solids has been barely explored and mostly using photons in the Hartley band, above 200 nm.[1,15,16] Still unanswered is the question of the photolysis of $O_2$ in the region of Schumann-Runge bands and continuum, where photoabsorption cross sections differ by orders of magnitude between the gas and the solid. Here we present results of experiments of the photolysis of optically thin (< 100 nm) solid films of pure ozone and oxygen at 22 K by 193 nm photons and a simple photochemical model from which we derive initial quantum yields, effective dissociation cross sections and probabilities for geminate recombination and reaction involving O atoms.

## 2. Experimental Setup

Experiments were performed in an ultrahigh vacuum chamber (base pressure ~ $10^{-10}$ Torr). Ozone was prepared by applying a Tesla coil discharge to the glass manifold pressurized with 100-150 Torr of $O_2$. Ozone formed from the discharge was condensed to the liquid phase by immersing the manifold in $LN_2$. After 1-2 minutes of discharge, we removed the remaining $O_2$ by pumping the manifold and admitted sublimating ozone molecules into the chamber through a PTFE valve.

Ozone molecules from the glass manifold were directed towards a gold-coated quartz crystal microbalance[17] cooled to 50 K. We chose this temperature to avoid condensation of trace $O_2$ from the glass manifold or $O_2$ produced from ozone reacting with any surface inside the vacuum chamber. The microbalance measures the areal mass of the condensed film, which is converted into column density (molecules $cm^{-2}$) by dividing by the molecular mass. We deposited films with column densities between 140 and 150 × $10^{15}$ $O_3$ $cm^{-2}$ at a rate that varied between 0.15 and 0.5 × $10^{15}$ $O_3$ $cm^{-2}$ $s^{-1}$. Using the density of 1.65 g $cm^{-3}$ for solid ozone[18,19], we obtain film thicknesses of ~67 - 73 nm. After deposition at 50 K, the ozone films were cooled to 22 K prior to UV exposure. To study solid $O_2$ we deposited films containing $230 \times 10^{15}$ $O_2$ $cm^{-2}$ at 22 K from a microcapillary-array doser at the rate of $0.3 \times 10^{15}$ $O_2$ $cm^{-2}$ $s^{-1}$. These films have a thickness of the 80 nm derived using a density of 1.53 g $cm^{-3}$ for solid oxygen.[20]

The films were irradiated with 193 nm pulses produced by ArF excimer laser at normal incidence. The laser beam was defocused using a $MgF_2$ lens ($f$ = 50 cm) to cover a rectangular spot ~ 25 mm × 75 mm at the target, larger than the active diameter of the microbalance (6 mm) to avoid multi-photon excitations or heating during each laser pulse. To measure the photon fluence reaching the target, we replaced the QCM with an Ophir power meter, which measured constant beam intensity across the area of the QCM. The laser delivered pulses of 0.84 mJ $cm^{-2}$ uniformly across the films at a rate that was varied from 0.2 to 5 Hz during the experiment. The energy per pulse was constant within 3% during the experiments. The values of fluence are calculated taking into account the ~20% light reflected by the gold substrate.[21]

The amount of ozone in the film was measured using specular reflectance infrared spectroscopy (35° incidence) at 2 $cm^{-1}$ resolution with a Thermo Nicolet Nexus 670 Fourier Transform Infrared Spectrometer. The $O_3$ column density was deduced from the area of the 9.5 μm absorption band due to the $\nu_3$ asymmetric O-O vibration in the ozone molecule, as explained below. The infrared spectra are expressed in optical depth units, -ln($R/R_0$), where $R/R_0$ is the ratio of the reflectance of the sample to the reflectance of the gold substrate.

The area of the $O_3$ band divided by an effective band strength $A*$ gives the ozone column density. The qualifier "effective" is used to emphasize that $A*$ is not a constant property of a molecule, but rather it depends on the thickness and the optical properties of the film.[22] Though the film thickness only changes slightly (~ 10%), the refractive index of the films, specially in the case of the pure ozone, changes significantly during 193 nm irradiation due to amorphization of the crystalline phase and the decrease in $O_3$ concentration in the film.[19]

For pure ozone, we obtain an initial effective band strength, $A_i^* = 4.7 \times 10^{-18}$ cm mol$^{-1}$ by dividing the $v_3$ band area by the $O_3$ column density measured by the microbalance following deposition. As the film is irradiated, ozone becomes diluted in $O_2$ and the band shape changes (Fig. 1). The final value $A_f^* = 1 \times 10^{-17}$ cm mol$^{-1}$, at the end of the experiment, is obtained using the desorption data of Figure 3. The same value is also obtained from desorption studies of irradiated $O_2$ films.

Obtaining the column density of ozone during photolysis of initially pure $O_2$ film is simple as the $v_3$ band shape does not change markedly with increasing fluence (Figure 4). This is expected since ozone is always diluted ($\leq 5$ %) in a matrix of solid oxygen. Thus we use $A^* = 1 \times 10^{-17}$ cm mol$^{-1}$ to derive ozone column densities for all fluences.

For ozone films, we approximate the two-fold variation of $A^*$ vs. fluence in the following manner. At low $F$, while the crystalline ozone amorphizes, we assume that $O_3$ is concurrently being destroyed with a phase-independent quantum yield, determined from the decrease in $O_3$ column density at higher $F$, discussed below. Therefore, below $3 \times 10^{16}$ photons cm$^{-2}$ there is a 13% ozone loss. However, the $v_3$ band area remained nearly constant despite irradiation. We assume that, due to the phase change, the value of $A^*$ increased to $5.4 \times 10^{-18}$ cm mol$^{-1}$. For large fluences, above $30 \times 10^{16}$ photons cm$^{-2}$ the band shape is essentially the same as that of ozone produced from irradiation of $O_2$ films, and therefore we use $A^* = 1 \times 10^{-17}$ cm mol$^{-1}$. For the intermediate region ($15 - 30 \times 10^{16}$ photons cm$^{-2}$), we use an average $A_a^* = 7.7 \times 10^{-18}$ cm mol$^{-1}$ to obtain column densities in this transitional regime. The uncertainty in the column density in the intermediate fluence region is 23 %.

## 3. Results

All 193 nm-photolysis experiments where performed at 22 K. Figure 1 shows changes in the $v_3$ band of a ~70 nm, initially pure, ozone film during photolysis. The band area decreases and the band position shifts to lower wavenumber with increasing photon fluence. Figure 2 shows that the amount of ozone present in the film decreases with fluence, nearly exponentially to a stationary level that is 5% of the initial $O_3$ column density.

In Figure 3, we compare the thermal desorption spectra of unprocessed and photolyzed ozone and oxygen films, while heating at 1 K min$^{-1}$. The desorption of the unprocessed ozone film peaks at 64 K while that of photolyzed $O_3$ has two peaks, at 32 K and a smaller peak at ~ 59 K . The desorption of unprocessed $O_2$ peaks at 32 K, while the photolyzed $O_2$ film also shows an additional smaller peak at 57 K. Photolysis converts most of $O_3$ to $O_2$, while photolysis of pure $O_2$ films produces small amount of $O_3$. After desorption of $O_2$ at 32 K the photolyzed films convert to low-density ozone[19] which desorbs at a somewhat lower temperature, with a peak at ~ 55- 60 K. The amount of ozone detected in photolyzed films of $O_2$ and $O_3$ by the microbalance gives us the final effective $O_3$ band strength $A_f* = 1 \times 10^{-17}$ cm mol$^{-1}$.

Figure 4 shows the appearance and growth of the $v_3$ ozone absorption band during photolysis of solid $O_2$. For large fluences, the band is remarkably similar to that measured on an $O_3$ film irradiated to $670 \times 10^{16}$ photons cm$^{-2}$, which confirms that films obtained at high fluences are essentially the same, when starting from pure $O_2$ or pure $O_3$ films. In Figure 5, we show a log-log plot of the fluence dependence of the production of $O_3$ in an irradiated $O_2$ film.

## 4. Discussion

To analyze our experiments we start by summarizing results of gas phase experiments and discuss differences between experiments done in the solid and with free molecules. In the gas phase, photoabsorption of a 193.3 nm (6.41 eV) photon dissociates $O_2$ into two ground state O ($^3$P) atoms. [24]

$$O_2 + h\upsilon \rightarrow 2\ O\ (^3P) \tag{1}$$

The absorption by a free molecule is in the Schumann-Runge region, where the cross section varies widely with wavelength due to extremely wide rovibrational oscillations. Thus, the results of measurements depend critically on gas temperature and pressure, and on the energy width of the exciting light, which may explain the range of reported cross sections[25]: 1.2 - 200 × 10$^{-22}$ cm$^2$. Due to the polarization of the medium and the intermolecular interactions, the energy levels in the solid are shifted and broadened compared with the gas phase, as seen in ultraviolet photoabsorption[1,26,27] and photoelectron emission[28] studies, with cross sections varying smoothly with photon energy. The photoabsorption cross section of solid $O_2$ at 193 nm is $\sigma_2 = 4.2 \times 10^{-19}$ cm$^2$, orders of magnitude larger than for the free molecule.[27]

For the free ozone molecule, photoabsorption leads to O ($^3$P) and electronically excited O ($^1$D) (1.97 eV) in roughly equal proportions[29,30], with a cross section of 4.3 × 10$^{-19}$ cm$^{-2}$.

$$O_3 + h\upsilon \rightarrow O\ (^1D) + O_2 \tag{2}$$

$$O_3 + h\upsilon \rightarrow O\ (^3P) + O_2 \tag{3}$$

where the $O_2$ product can result in different excited states. Dissociation into three O atoms[30], or production of O ($^1$S) [Ref.[31]] have yields two orders of magnitude smaller than for O ($^1$D) or O($^3$P) production. In the solid state, photoabsorption cross section of ozone at 193 nm is $\sigma_3 = 2.5$ $\times 10^{-18}$ cm$^2$, obtained by extrapolating measurements of Sedlacek and Wight[1] from 200 to 193 nm. This value is 5.8 times larger than for gas phase ozone.

In the gas phase, the main reactions of the oxygen radicals are:

$$O\,(^1D) + O_2 + M \rightarrow O_3 + M \qquad\qquad (4)$$

$$O\,(^3P) + O_2 + M \rightarrow O_3 + M \qquad\qquad (5)$$

$$O\,(^1D) + O_3 \rightarrow 2\,O_2 \qquad\qquad (6)$$

$$O\,(^3P) + O_3 \rightarrow 2\,O_2 \qquad\qquad (7)$$

$$O_2\,(X\,^3\Sigma_g^-,\,v^*) + O_2 \rightarrow O_3 + O \qquad\qquad (8)$$

where $v^*$ is a high vibrational excitation and M is a third body needed to conserve energy and momentum in the association reactions.

In the solid state, the picture is modified not only because of possibly different branching ratios, but also because of the high frequency of collisions that produce many ways to dissipate kinetic, vibrational and electronic energy. Prominent is the cage effect, where dissociation products can bounce back from neighboring molecules (the cage) and recombine (geminate recombination), thereby lowering the effective dissociation cross section.

The high molecular density in the solid-state has multiple implications. For instance, a photon may actually interact with a pair of $O_2$ molecules, or dimer, a process that causes absorption in the visible region in the Earth's atmosphere and on planetary ices[32] and also causes ultraviolet absorption at photon energies between 4.5 and 7 eV in $O_2$ at high pressures[33]. Photoabsorption by the $(O_2)_2$ dimer is much more efficient than the dipole-forbidden transitions of the isolated molecule, and is likely the reason for the larger photoabsorption cross section of the solid at 193 nm mentioned above.

The transformation of electronic energy into kinetic energy during dissociation is altered as the recoiling products in the solid interact with surrounding molecules, setting them in motion as well. This sharing of kinetic energy implies that the energy of a given dissociation product, say, an O atom, is lower than in the gas phase and cannot be derived simply from the photon energy and the bond energy of the molecule. In the following, we will assume that reaction products dissipate their energy (vibrational, electronic, and kinetic) before they react, because of many body interactions in the solid, including multi-phonon relaxation processes and multiple crossing of potential energy surfaces. We will present below a simple photochemical model based on the assumptions which describes the data well, except for small differences that may be attributed to incomplete energy relaxation.

With this simplification, we consider two initial processes:

$$O_2 + h\nu \rightarrow 2\,O \tag{9}$$

$$O_3 + h\nu \rightarrow O + O_2 \tag{10}$$

and three subsequent reactions:

$$O + O \rightarrow O_2 \tag{11}$$

$$O + O_2 \rightarrow O_3 \tag{12}$$

$$O + O_3 \rightarrow O_2 + O_2 \tag{13}$$

Here, conservation of energy and momentum in reactions (9) and (10) is allowed by the presence of the lattice.

*4.1. Initial cross sections and quantum yields in solid ozone*

For very low fluences, the relative variation of the ozone column density $\eta_3$ with $F$ is given by:

$$\frac{d\eta_3}{dF} = -m_3(1-\alpha_3)\sigma_3\eta_3 \qquad (14)$$

which integrates to

$$\ln\left[\frac{\eta_3}{\eta_3(0)}\right] = -m_3(1-\alpha_3)\sigma_3 F . \qquad (15)$$

Here $\sigma_3$ is the photoabsorption cross section and $\alpha_3$ is the probability of geminate recombination of O + $O_2$ survives recombination. The factor $m_3$ takes into account the total number of ozone molecules destroyed from absorption of a photon and subsequent reactions induced by the O photodissociation fragments. The product $m_3(1-\alpha_3)$ gives the initial quantum yield, or number of ozone molecules destroyed per absorbed photon and $\sigma_3' = (1-\alpha_3)\sigma_3$ is the net dissociation cross section in the solid.

Figure 6 shows a fit of Eq. 15 to the experimental $\eta_3(F)$, which gives a slope $m_3(1-\alpha_3)\sigma_3$ = $(4.3 \pm 0.4) \times 10^{-18}$ cm$^2$. This is nearly twice $\sigma_3$ [Ref. 1], i.e., the initial quantum yield for $O_3$ photodestruction is $\Phi(-O_3)$ =1.7 $\pm$ 0.1, similar to the value of 1.5 $\pm$ 0.2 for 308 nm photons, on the higher wavelength edge of the Hartley band.[1] Since another ozone molecule is destroyed, according to Eq. 13, after each dissociation event, $m_3 \sim 2$. Therefore, the probability for geminate recombination of $O_2$ and O is $\alpha_3$ = 0.15.

*4.2. Initial cross sections and quantum yields in solid $O_2$*

The O atoms resulting from photoabsorption and subsequent dissociation can either recombine to reform $O_2$ or react with additional $O_2$ to form two ozone molecules: $-3\ O_2 \rightarrow 2\ O_3$. The net photodissociation of $O_2$ leading to the formation of $O_3$ in an oxygen matrix is described by

$$\frac{d\eta_2}{dF} = -m_2(1-\alpha_2)\sigma_2\eta_2 \tag{16}$$

where $m_2$ is the number of $O_2$ molecules that are destroyed per net dissociation event. Since the number of oxygen atoms in the matrix is conserved, Eq. 16 is integrated to:

$$\ln\left[1 - \frac{3\eta_3}{2\eta_2(0)}\right] = -m_2(1-\alpha_2)\sigma_2 F \tag{17}$$

Figure 6 shows a fit of Eq. 17 to the data for solid $O_2$ at low fluences), which gives a slope of $m_2(1-\alpha_2)\sigma_2 = (1.19 \pm 0.08)$ x $10^{-19}$ cm$^2$. Since $\sigma_2 = 4.2$ x $10^{-19}$ cm$^2$ (Ref. [27,28]), the quantum yield for $O_2$ destruction is $\Phi(-O_2) = 0.28 \pm 0.02$. Since each dissociation of $O_2$ produces two oxygen atoms which in turn react with two additional $O_2$ molecules producing $O_3$, $m_2 = 3$. Thus, the probability of geminate recombination is $\alpha_2 = 0.90$, six times the value for ozone. Each dissociation produces two ozone molecules, and therefore $\Phi(O_3) = 0.18 \pm 0.01$ for ozone synthesis.

*4.3. Rate equations and fluence dependence*

Modeling ozone production and destruction on the basis of the Chapman equations is difficult because of the existence of two widely different time scales[34]: those for reactions of atomic O atoms and for the buildup of radiation products in the film. We circumvent this problem by assuming that O atoms are consumed by reactions as soon as they are formed. This means that they do not accumulate in the film to a significant level, which should be a good assumption based on previous discussions.[8] Thus, the evolution of the $O_3$ and $O_2$, determined by a competition between creation and destruction rates, is given by:

$$\frac{d\eta_3}{dF} = -\sigma_3^{'}\eta_3 - \sigma_3^{'}\eta_3\frac{\eta_3}{\eta*} + \sigma_3^{'}\eta_3 R_{23}\frac{\eta_2}{\eta*} - 2\sigma_2^{'}\eta_2\frac{\eta_3}{\eta*} + 2\sigma_2^{'}\eta_2 R_{23}\frac{\eta_2}{\eta*} \qquad (18)$$

The first term accounts for reduction in $O_3$ column density due to photoabsorption followed by dissociation, which produces O and $O_2$ fragments; the other terms describe reactions of O atoms with additional $O_2$ or $O_3$ molecules which occur according to their relative abundance in the films and the respective reaction cross sections. The second and the third terms are for O formed from photodissociation of $O_3$; the last two are for the two O atoms from photodissociation of $O_2$.

Equation 18 reduces to

$$\frac{d\eta_3}{dF} = -\sigma_3^{'}\eta_3 - \frac{\eta_3}{\eta*}\left(\sigma_3^{'}\eta_3 + 2\sigma_2^{'}\eta_2\right) + R_{23}\frac{\eta_2}{\eta*}\left(\sigma_3^{'}\eta_3 + 2\sigma_2^{'}\eta_2\right) \qquad (19)$$

where $R_{23}$ is the ratio of reaction cross sections of O with $O_2$ and $O_3$ and $\eta* = R_{23}\,\eta_2 + \eta_3$.

Eq. 19 has only one unknown, $R_{23}$, and can be used to fit the numerical derivatives of the fluence dependences of $O_3$ column density in the two experiments, as shown in Figure 7. The fit to both experiments gives $R_{23} = 0.10 \pm 0.01$, meaning that O atoms are 10 times less likely to react when they encounter an $O_2$ vs. an $O_3$ molecule. Remarkably, we obtain the same value of $R_{23}$, within errors, for both $O_2$ and $O_3$ films, even though the energy of oxygen atoms resulting from $O_3$ dissociation is much larger than from $O_2$ dissociation since the $O_2$-O bond energy of 1 eV being much smaller than that for O-O, 5 eV. The results then suggest that the O atoms can rapidly lose their excess kinetic energy through multiple relaxation pathways available in the solid prior to their subsequent reactions. The relaxation pathways should also be available for rapid quenching of the vibrationally excited $O_2$, and justifies its neglect in the model.

The solid lines in Figure 2 and 5 are integrals of the fits in Figure 7. Table I summarize the results of these experiments and compare them with gas phase values. The remarkably higher photoabsorption cross section of solid $O_2$ at 193 nm in the solid, vs. the gas, is the cause for higher $\eta_3 : \eta_2$ ratio in at the end of irradiation in solid phase. This is likely also the reason for the observed 50 fold increase in the final $\eta_3 : \eta_2$ ratio observed in photolysis of solid $O_2$ at 210 nm compared to 250 nm[16].

## 5. Astrophysical Implications

The prospect of the presence of reactive species like ozone in low temperature environments like the molecular clouds is interesting because radiolytic O atoms can start multiple routes of molecular synthesis in the interstellar medium. An estimate of the solid $O_2$ abundance in molecular clouds[13], based on observations of the weak $O_2$ transition at 6.45 μm and the profile of the 4.67 μm CO absorption feature, gave upper limits of $6 - 12 \times 10^{17}$ $O_2$ cm$^{-2}$, which only accounts for ~ 6% of total oxygen budget in the ISM. These $O_2$ condensed on the grains can be partially converted into ozone by energetic cosmic rays[35] or the UV radiation they generate in the background gas[36]. Though, so far, results of observations have been negative on the presence of $O_3$ on the grains, it is possible that the 9.6 μm ozone absorption is masked by a broad absorption feature at ~10 μm due to silicates[11] or other materials.[37] As we have demonstrated, ozone is dissociated by UV photons and its hot reactive atoms could react with other molecules, such as CO to produce $CO_2$. If this process is relevant, it could help resolve the problem of accounting for the observed abundance of solid $CO_2$ in interstellar clouds.[38]

## Acknowledgment

This work was supported by NASA Grants NNX07AL48G (Cosmochemistry) and NNX08AI44G (Origins of the Solar System).

## References

[1] A. J. Sedlacek and C. A. Wight, The Journal of Physical Chemistry **93** (2), 509 (1989).

[2] V. B. Alexander and A. W. Charles, The Journal of Chemical Physics **101** (1), 292 (1994).

[3] K. S. Noll, R. E. Johnson, A. L. Lane, D. L. Domingue, and H. A. Weaver, Science **273** (5273), 341 (1996); K. S. Noll, T. L. Roush, D. P. Cruikshank, R. E. Johnson, and Y. J. Pendleton, Nature **388**, 45 (1997).

[4] D. A. Bahr, M. Fama, R. A. Vidal, and R. A. Baragiola, J. Geophys. Res. **106**, 33285 (2001); G. P. Nikolay, G. K. Alexander, and A. K. Greg, The Journal of Chemical Physics **125** (12), 124702 (2006).

[5] B. D. Teolis, M. J. Loeffler, U. Raut, M. Fama, and R. A. Baragiola, The Astrophysical Journal Letters **644** (2), L141 (2006).

[6] M. Famá, D. A. Bahr, B. D. Teolis, and R. A. Baragiola, Nuclear Instruments and Methods in Physics Research B **193**, 775 (2002).

[7] S. Lacombe, F. Cemic, K. Jacobi, M. N. Hedhili, Y. Le Coat, R. Azria, and M. Tronc, Physical Review Letters **79**, 1146 (1997).

[8]     R. A. Baragiola, C. L. Atteberry, D. A. Bahr, and M. M. Jakas, Nuclear Instruments & Methods in Physics Research Section B-Beam Interactions with Materials and Atoms **157** (1-4), 233 (1999).

[9]     P. D. Cooper, M. H. Moore, and R. L. Hudson, Icarus **194** (1), 379 (2008); C. J. Bennett and R. I. Kaiser, Astrophysical Journal **635**, 1362 (2005).

[10]    P. A. Gerakines, W. A. Schutte, and P. Ehrenfreund, Astronomy and Astrophysics **312**, 289 (1996).

[11]    P. Ehrenfreund, M. Burgdorf, L. D'Hendecourt, and J. M. Greenberg, Advances in Space Research **13**, 465 (1993).

[12]    A. G. G. M. Tielens and W. Hagen, Astronomy and Astrophysics **114**, 245 (1982); P. Ehrenfreund, R. Breukers, L. D'Hendecourt, and J. M. Greenberg, Astronomy and Astrophysics **260**, 431 (1992).

[13]    B. Vandenbussche, P. Ehrenfreund, A. C. A. Boogert, E. F. van Dishoeck, W. A. Schutte, P. A. Gerakines, J. Chiar, A. G. G. M. Tielens, J. Keane, D. C. B. Whittet, M. Breitfellner, and M. Burgdorf, Astronomy and Astrophysics **346**, L57 (1999).

[14]    D. H. Parker, Accounts of Chemical Research **33** (8), 563 (2000); T. G. Slanger and R. A. Copeland, Chemical Reviews **103** (12), 4731 (2003); Y. Matsumi and M. Kawasaki, Chemical Reviews **103** (12), 4767 (2003).

[15]    H. Chaabouni, L. Schriver-Mazzuoli, and A. Schriver, Low Temperature Physics **26** (9), 712 (2000).

[16]    M. J. Dyer, C. G. Bressler, and R. A. Copeland, Chemical Physics Letters **266** (5-6), 548 (1997).

[17]    N. J. Sack and R. A. Baragiola, Physical Review B **48** (14), 9973 (1993).

[18]    A. G. Streng and A. V. Grosse, Journal of the American Chemical Society **81** (4), 805 (1959).

[19]    B. D. Teolis, M. Famá, and R. A. Baragiola, Journal of Chemical Physics **127**, 4507 (2007).

[20]    C. S. Barrett and L. Meyer, Physical Review **160** (3), 694 (1967).

[21]    L. R. Canfield, G. Hass, and W. R. Hunter, Journal de Physique **25** (1-2), 124 (1964).

[22]    B. D. Teolis, M. J. Loeffler, U. Raut, M. Famá, and R. A. Baragiola, Icarus **190** (1), 274 (2007).

[23]    M. P. Collings, M. A. Anderson, R. Chen, J. W. Dever, S. Viti, D. A. Williams, and M. R. S. McCoustra, Monthly Notices of the Royal Astronomical Society **354**, 1133 (2004).

[24]    H. Okabe, *Photochemistry of small molecules*. (Wiley, New York, 1978).

[25]    H. T. Liou, S. F. Chiou, and K. L. Huang, Ozone: Science & Engineering **19** (3), 273 (1997); Y. Niwa, A. Matsuzaki, S. Nishio, H. Sato, and I. Tanaka, The Journal of Physical Chemistry A **101** (4), 668 (1997); K. Yoshino, D. E. Freeman, J. R. Esmond, and W. H. Parkinson, Planetary and Space Science **31** (3), 339 (1983).

[26]    S. R. Scharber, Jr. and S. E. Webber, The Journal of Chemical Physics **55** (8), 3977 (1971); N. J. Mason, A. Dawes, P. D. Holtom, R. J. Mukerji, M. P. Davis, B. Sivaraman, R. I. Kaiser, S. V. Hoffmann, and D. A. Shaw, Faraday Discussions **133**, 311 (2006).

[27]    H.-C. Lu, H.-K. Chen, B.-M. Cheng, and J. F. Ogilvie, Spectrochimica Acta Part A: Molecular and Biomolecular Spectroscopy **71** (4), 1485 (2008).

[28]  F.-j. Himpsel, N. Schwentner, and E. E. Koch, physica status solidi (b) **71** (2), 615 (1975).

[29]  A. T. Andrew, L. V. Ghanshyam, G. Tomasz, E. T. John, and A. R. Ravishankara, The Journal of Chemical Physics **95** (5), 3244 (1991).

[30]  S. Nishida, F. Taketani, K. Takahashi, and Y. Matsumi, The Journal of Physical Chemistry A **108** (14), 2710 (2004).

[31]  K. Takahashi, T. Nakayama, and Y. Matsumi, The Journal of Physical Chemistry A **107** (44), 9368 (2003).

[32]  R. A. Vidal, D. Bahr, R. A. Baragiola, and M. Peters, Science **276** (5320), 1839 (1997).

[33]  Y. Akahama and H. Kawamura, Chemical Physics Letters **392** (4-6), 476 (2004).

[34]  R. M. Velasco, F. J. Uribe, and E. Pérez-Chavela, Journal of Mathematical Chemistry **44** (2), 529 (2008).

[35]  A. G. G. M. Tielens, *The Physics and Chemistry of the Interstellar Medium*. (2005).

[36]  S. S. Prasad and S. P. Tarafdar, Astrophysical Journal **267**, 603 (1983).

[37]  J. H. Lacy, H. Faraji, S. A. Sandford, and L. J. Allamandola, The Astrophysical Journal Letters **501** (1), L105 (1998).

[38]  V. Mennella, M. E. Palumbo, and G. A. Baratta, Astrophysical Journal **615**, 1073 (2004).

**Figures**

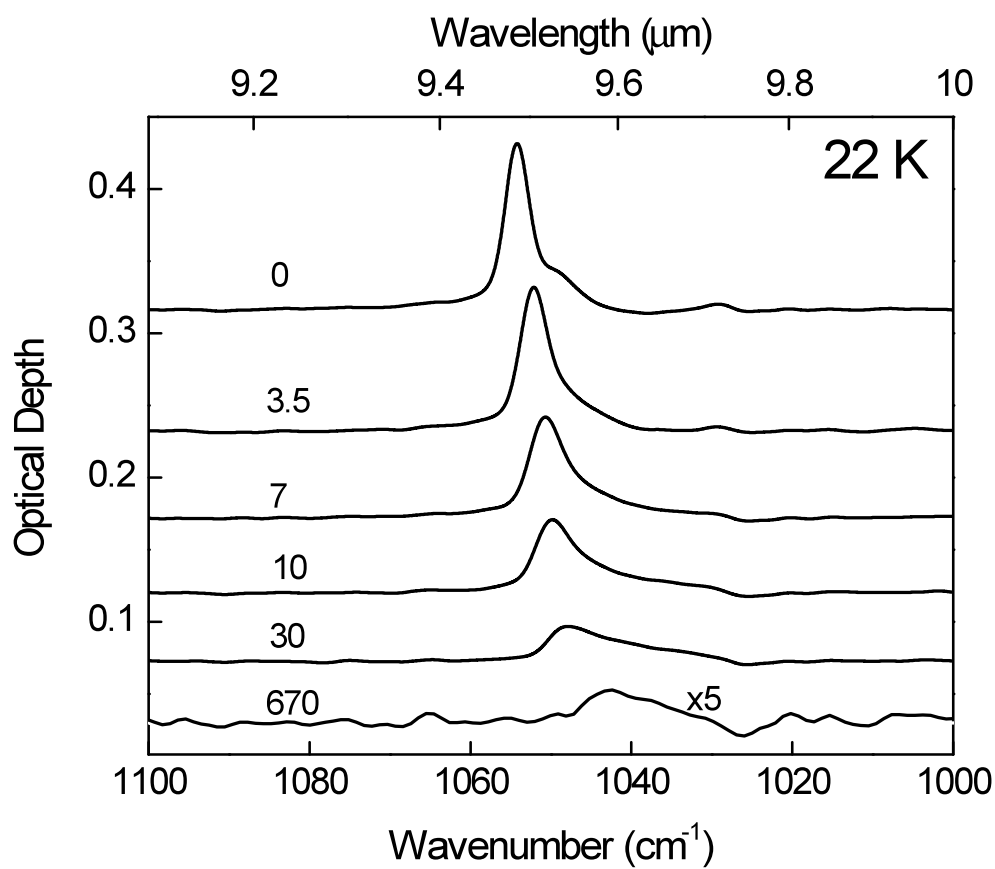

Figure 1. Optical depth in the region of the $\upsilon_3$ ozone absorption showing a decrease in band area following exposure to 193 nm photons. The ozone film ($146 \times 10^{15}$ $O_3$ cm$^{-2}$) was deposited at 50

K and cooled to 22 K prior to photolysis. The numbers displayed with the spectra correspond to the fluence in units of $10^{16}$ photons cm$^{-2}$. The spectra have been displaced vertically for clarity.

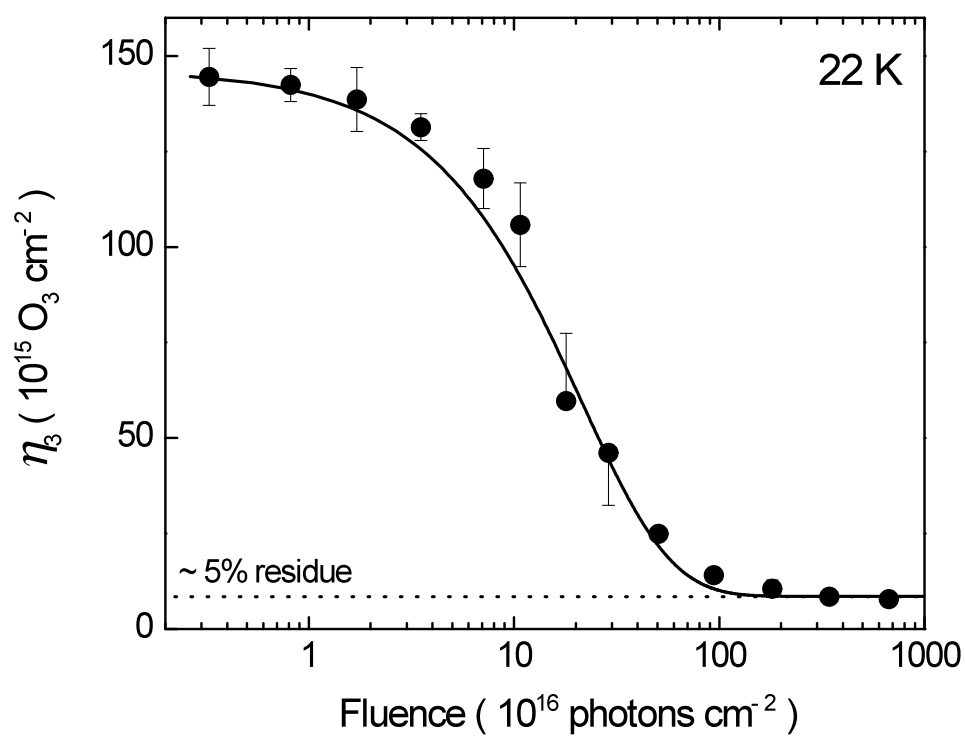

Figure 2. Decrease in the $O_3$ column density estimated from 9.5 μm $v_3$ ozone absorption band as the ozone film is irradiated with 193 nm light. A residual amount of $O_3$ (5% of the initial value) remains in the film at the end of the photolysis. The solid line is the integral of fit (Figure 7) obtained from a model that describes the rate of $O_3$ destruction from photolysis of solid $O_3$ film. The details of the model are in the text.

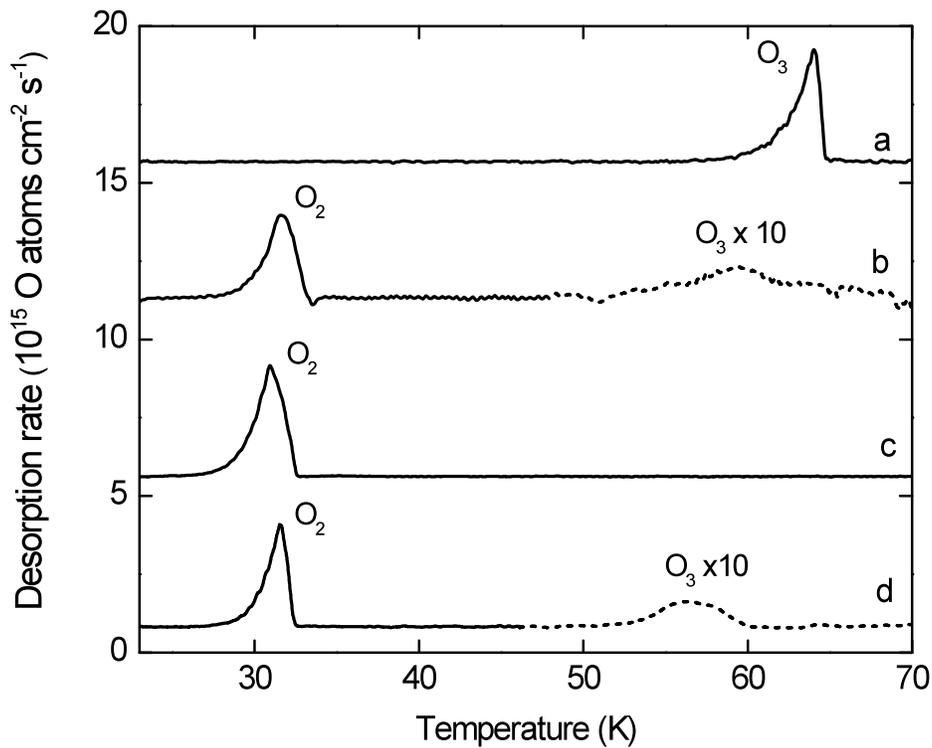

Figure 3. Thermal desorption spectra at 1 K min$^{-1}$ of (a) unprocessed ozone film ($141 \times 10^{15}$ O$_3$ cm$^{-2}$), (b) photolyzed ozone film ($146 \times 10^{15}$ O$_3$ cm$^{-2}$), (c) unprocessed O$_2$ film ($230 \times 10^{15}$ O$_2$ cm$^{-2}$) and (d) photolyzed O$_2$ film ($230 \times 10^{15}$ O$_2$ cm$^{-2}$). The desorption rates were obtained from the time derivative of the mass loss measured with the microbalance. The dotted sections are multiplied by 10 to show the small ozone desorption peak.

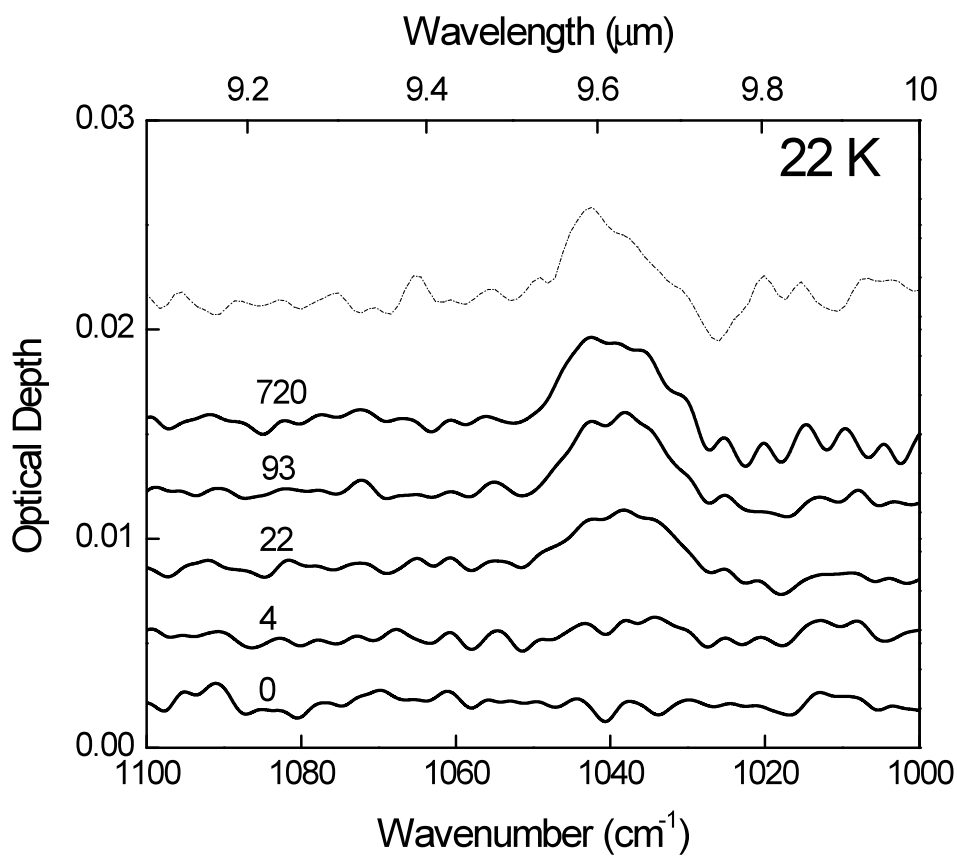

Figure 4. Optical depth in the region of the $\upsilon_3$ ozone absorption for a solid $O_2$ film ($\sim 230 \times 10^{15}$ $O_2$ cm$^{-2}$) irradiated at 22 K, showing the increase in the band area with photon fluence, in units of $10^{16}$ photons cm$^{-2}$. The top dashed curve is for initially pure ozone film irradiated to $670 \times 10^{16}$ photons cm$^{-2}$. Its similarity with the spectrum of the pure $O_2$ irradiated to highest fluence ($720 \times 10^{16}$ photons cm$^{-2}$) confirms that the initially pure $O_2$ or $O_3$ films approach the same steady state concentrations at high fluences. The spectra have been displaced vertically for clarity.

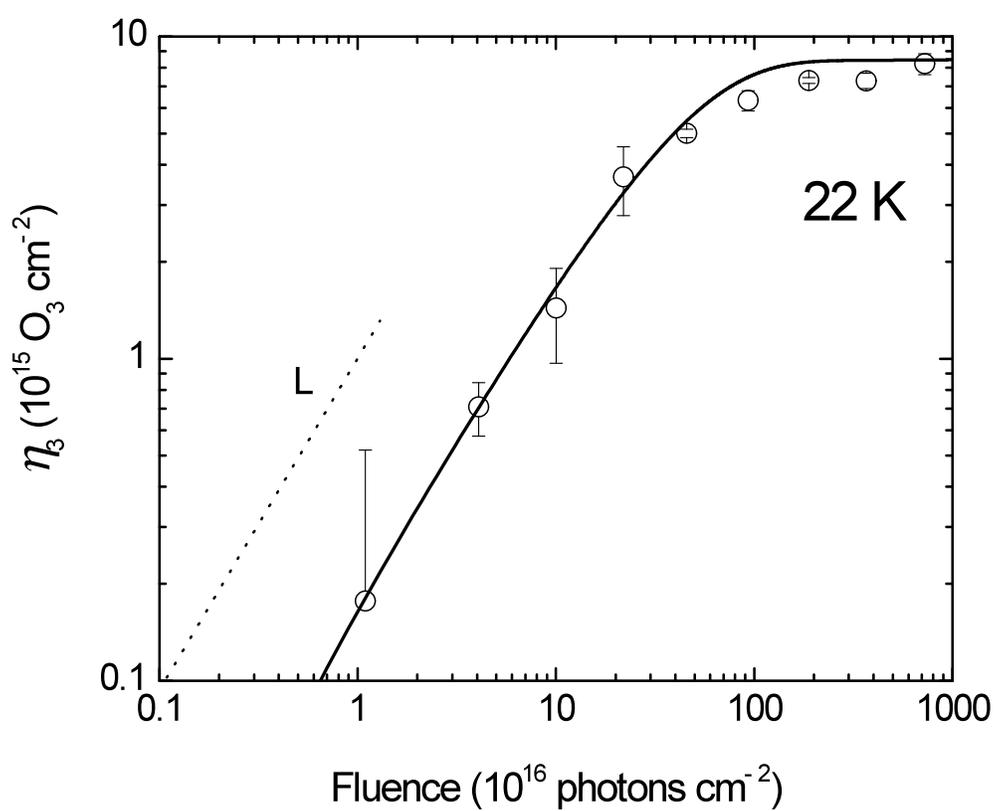

Figure 5. Fluence dependence of $O_3$ production from irradiation of a solid $O_2$ film with 193 nm light at 22 K. The solid line is the integral of the fit (Figure 7) obtained from the analytical model (see text). The dotted line labeled L (Linear) emphasizes the initial linear fluence dependence.

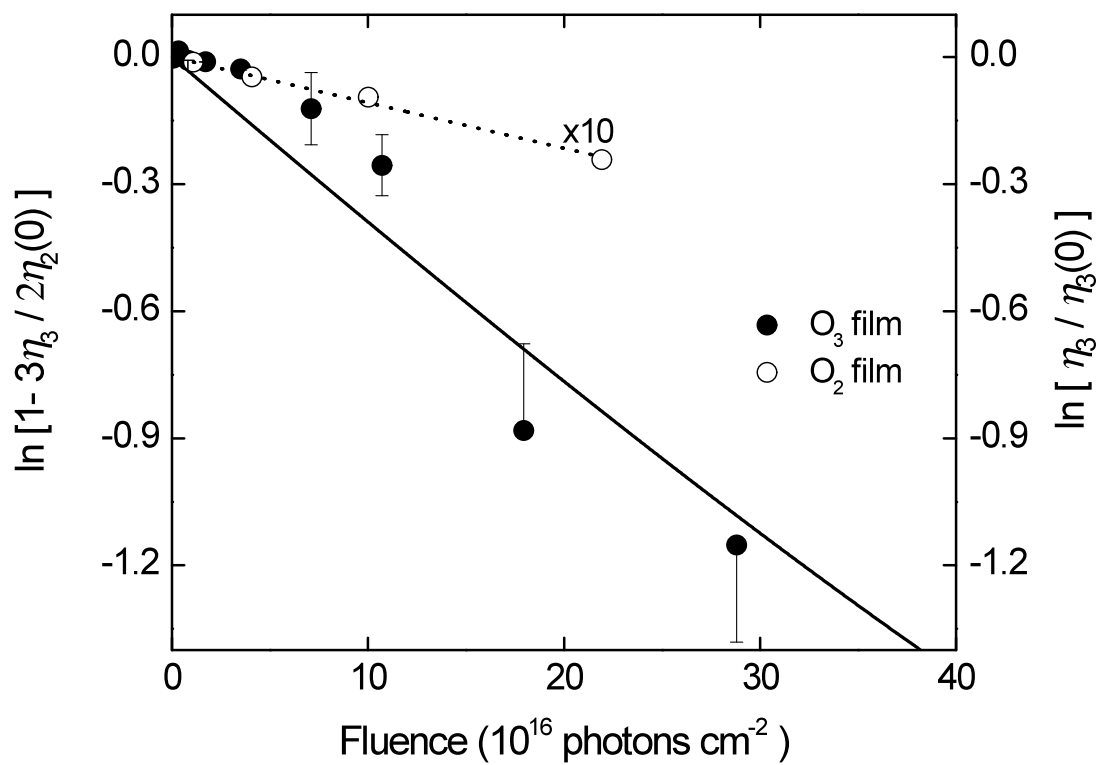

Figure 6. Logarithmic representation of the initial $O_3$ destruction (adapted from Figure 2) and $O_3$ synthesis (adapted from Figure 5). The slope of the solid line is equal to the $m_3(1-\alpha_3)\sigma_3$, where $m_3(1-\alpha_3)$ is the initial $O_3$ photodestruction quantum yield and $\sigma_3$ is the photoabsorption cross section of $O_3$. $\alpha_3$ is the probability of geminate recombination for $O_2$ and $O$. The slope of the dotted line is equal to the $m_2(1-\alpha_2)\sigma_2$, where $m_2(1-\alpha_2)$ is the initial quantum yield of $O_2$ photoproduction and $\sigma_2$ is the photoabsorption cross section of solid $O_2$. $\alpha_2$ is the probability for geminate recombination of the two $O$ atoms from dissociation of $O_2$.

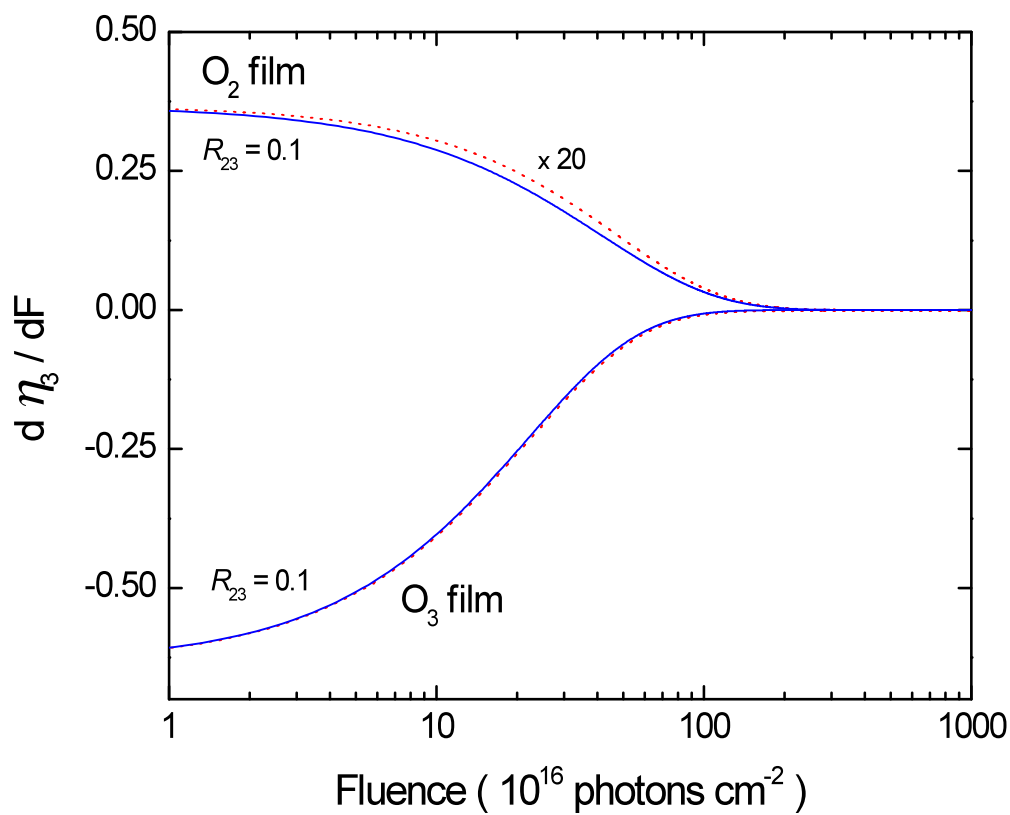

Figure 7. The rate of change in the $O_3$ column density vs. fluence, $d\eta_3/dF$. The red dotted lines are the numerical derivative of fits to the experimental $\eta_3(F)$ in photolysis of solid $O_2$ and $O_3$. The blue solid lines are fits obtained from the model discussed in the text, by optimizing the ratio of reaction cross sections $R_{23}$. The fits give $R_{23} = 0.10 \pm 0.01$.

Table I. Comparison of different parameters obtained from 193 nm photolysis of solid $O_2$ and $O_3$ film with gas phase values

| | Solid | Gas (from literature) |
|---|---|---|
| $\sigma_2{'}$ (cm$^2$) | $(3.96 \pm 0.04) \times 10^{-20}$ | $1.2 - 200 \times 10^{-22}$ [a] |
| $\sigma_3{'}$ ($10^{-18}$ cm$^2$) | $2.1 \pm 0.2$ | $0.43$ [b] |
| $\alpha_3$ | $0.15$ | - |
| $\alpha_2$ | $0.90$ | - |
| $\Phi(-O_3)$ | $1.7 \pm 0.1$ | |
| $\Phi(-O_2)$ | $0.28 \pm 0.02$ | $0.45*$ (300- 1300 Torr) [c] |
| $\Phi(O_3)$ | $0.19 \pm 0.01$ | $0.3$ (300- 1300 Torr) [d] |
| $\eta_3/\eta_2 (F \rightarrow \infty)$ | $\sim 0.03$ | $0.001 - 0.003$ (50-800 Torr) [e] |
| $R_{23}$ | $0.10 \pm 0.01$ | |

(a) *Liou et. al.*, Ozone: Science and Engineering, 19:3, 273, 1997; *Niwa et. al.*, The Journal of Physical Chemistry A, 101, 4, 668, 1997; *Yoshino et. al.*, Planetary and Space Science, 31, 3, 339, 1983

(b) *Molina et. al.*, Journal of Geophysical Research, 91, D13, 14501, 1986

(c, d) *Washida et.al.*, The Journal of Chemical Physics, 54, 3, 1119, 1971

(e) *Liou et. al.*, Ozone: Science and Engineering, 19:3, 273, 1997